\documentclass[12pt]{article}
\usepackage{color}
\newcommand{\beq}{\begin{equation}}
\newcommand{\eeq}{\end{equation}}
\newcommand{\beqa}{\begin{eqnarray}}
\newcommand{\eeqa}{\end{eqnarray}}
\newcommand{\beqar}{\begin{eqnarray*}}
\newcommand{\eeqar}{\end{eqnarray*}}

\newcommand{\al}{\alpha}
\newcommand{\be}{\beta}

\def\spa          {\ \ \ }
\def\non          {\nonumber}
\def\ha           {\mbox{$\frac{1}{2}$}}

\def\spa          {\ \ \ }
\def\mand         {\spa\mbox{and}\spa}

\def\Tr           {\mbox{\rm Tr}\,}
\def\STr          {\mbox{\rm STr}\,}
\def\Str          {\mbox{\rm Str}\,}

\def\cd           {{\cdot}}
\def\ran          {\rangle}
\def\lan          {\langle}

\def\fsH    {H\!\!\!\!/\,}
\newcommand{\del}{\delta}

\newcommand{\eps}{\epsilon}
\newcommand{\ga}{\gamma}

\newcommand{\inn}{\!\cdot\!}

\newcommand{\lam}{\lambda}

\newcommand{\sig}{\sigma}

\newcommand{\z}{\zeta}

\newcommand{\ie}{{\it i.e.,}\ }
\newcommand{\labell}[1]{\label{#1}} %{\label{#1}} %
\newcommand{\reef}[1]{(\ref{#1})}
\newcommand\prt{\partial}

\usepackage{amsmath}
\usepackage{slashed}

\def\sst#1{{\scriptscriptstyle #1}}
\def\0{{\sst{(0)}}}
\def\1{{\sst{(1)}}}
\def\2{{\sst{(2)}}}
\def\3{{\sst{(3)}}}
\def\4{{\sst{(4)}}}
\def\5{{\sst{(5)}}}
\def\6{{\sst{(6)}}}
\def\7{{\sst{(7)}}}
\def\8{{\sst{(8)}}}

\begin{document}
\baselineskip 18pt%
\begin{titlepage}
\vspace*{1mm}%
\hfill
\vbox{

    \halign{#\hfil         \cr
   %       hep-th/yymmnnn\cr
       %  ICTP-PH-TH/2012-xyz\cr
         %IPM/P-2010/003  \cr
         %  CPHT RR-xxx .yyzz \cr
           } % end of \halign
      }  % end of \vbox
\vspace*{9mm}
\vspace*{9mm}%

\center{{\bf\Large On  RR Couplings and Bulk Singularity Structures of Non-BPS Branes 
%in Type II Superstring Theory
}}\vspace*{1mm} \centerline{{\Large {\bf  }}}
%\vspace*{1mm}
\begin{center}
{Ehsan Hatefi   \small $^{1,2,3}$}

\vspace*{0.3cm}
%{ {\it %{\small $^{1}$International Centre for Theoretical Physics, Strada Costiera 11, Trieste, Italy},
\vskip.1in
{ $^{1}$ Centre for Research in String Theory, School of Physics and Astronomy,
\\
Queen Mary University of London, Mile End Road, London E1 4NS, United Kingdom},
\vskip.06in
%,\\and
{ $^{2}$ Institute for Theoretical Physics, TU Wien
\\
Wiedner Hauptstrasse 8-10/136, A-1040 Vienna, Austria}
%\vskip.06in
%{ $^{3}$ Institute des Hautes Etudes Scientifiques Bures-sur-Yvette, F-91440, France }
%\footnote{ehsanhatefi@gmail.com, }
\vskip.06in
{ $^{3}$ e.hatefi@qmul.ac.uk, ehsan.hatefi@tuwien.ac.at, ehsan.hatefi@cern.ch}
\vspace*{0.1cm}
\vspace*{.1cm}
\end{center}
\begin{center}{\bf Abstract}\end{center}
\begin{quote}

We compute the five point world sheet scattering amplitude of a symmetric closed string Ramond-Ramond , a transverse scalar field, a world volume gauge field and a real tachyon in both world volume and transverse directions of brane in type IIA and IIB superstring theory. We provide the complete analysis of $<C^{-1} \phi^0  A^0 T ^{-1}>$ S-matrix and show that both $u'=u+\frac{1}{4}$ and $t$  channel bulk singularity structures can also be examined by this S-matrix.  Various remarks about new restricted Bianchi identities on world volume for the other pictures have also been made.

\end{quote}
\end{titlepage}%--------------------------------------------------------------------

\section{Introduction}

No matter we talk about stable BPS or unstable (non-BPS) branes,  $D_{p}$-branes  are supposed to be thought of  sources for closed string Ramond-Ramond (RR) field   \cite{Polchinski:1995mt}. It has been proven that making use of RR couplings, one can investigate or try to address various remarkable issues about string theory, whereas we highlight some of the most crucial ones  as follows.  The  so called  brane embeddings  \cite{Douglas:1995bn}, through D-brane language the K-theory \cite{Witten:1998cd}, also the so called Dielectric effect or  Myers effect  \cite{Myers:1999ps} and in particular the way of looking for all order $\alpha'$ higher derivative corrections to BPS or non-BPS branes \cite{Hatefi:2012zh},\cite{Hatefi:2013mwa} are explored. Note that for the definitions of non-BPS branes, $D_{p}$-branes with  $p$  odd (even) in IIA (IIB) are taken in which  $p$ stands for the spatial dimension of branes.  Based on various symmetries, a universal conjecture for  all order   $\alpha'$ higher derivative corrections to both BPS and non-BPS branes was proposed in  \cite{Hatefi:2012rx} and naturally it has been applied at practical levels to various higher point fermionic S-matrices \cite{Hatefi:2013hca} as well.

 \vskip .1in
 
 It  was  argued in \cite{Sen:2004nf} in detail that, to deal with effective theory of unstable branes after integrating out all the massive modes, one is just left with massless and tachyon states , and also to work out the dynamics of branes not only DBI  action but also Wess-Zumino terms are indeed needed. To obtain effective actions, either Boundary String Field theory (BSFT) \cite{Kraus:2000nj} or S-matrix formalism should be employed where the latter has a very strong potential to be taken so that upon applying that,   all the coefficients of higher derivative corrections  for all orders in $\alpha'$ can be found.

  The  Wess-Zumino effective action has been given in \cite{Hatefi:2012wj} as 
\beqa
S_{WZ}&=&\mu_p' \int_{\Sigma_{(p+1)}} C \wedge \Str e^{i2\pi\alpha'\cal F}\labell{WZ'},\eeqa
with   $\beta'$ being normalisation  constant and with the so called the curvature of super connection as follows
\beqa
i{\cal F} = \left(
\begin{array}{cc}
iF -\beta'^2 T^2 & \beta' DT \\
\beta' DT & iF -\beta'^2T^2
\end{array}
\right) \label{WZ44},\eeqa
Now if we expand the exponential, one produces various couplings such as 
\beqa
S_{WZ}=2\beta'\mu_p' (2\pi\alpha')\Tr\left(C_{p}\wedge DT + (2\pi\alpha')C_{p-2}\wedge DT\wedge F+
\frac{(2\pi\alpha')^2}{2}C_{p-4}\wedge F\wedge F\wedge DT\right)\label{WZ66}\eeqa

Based on the internal Chan-Paton  matrix some partial selection rules for superstring amplitudes have been released in \cite{Hatefi:2013yxa}, and to make sense of S-matrix computations, one needs to keep track of  internal CP matrix of tachyons around unstable point of tachyon DBI action. The final form of tachyon DBI up to some orders with its all  ingredients is demonstrated in \cite{Hatefi:2012wj} to be as follows
\beqa
S_{DBI}&\sim&\int
d^{p+1}\sigma \STr\left(\frac{}{}V({ T^iT^i})\sqrt{1+\frac{1}{2}[T^i,T^j][T^j,T^i])}\right.\labell{nonab} \\
&&\qquad\qquad\left.
\times\sqrt{-\det(\eta_{ab}
+2\pi\alpha'F_{ab}+2\pi\alpha'D_a{ T^i}(Q^{-1})^{ij}D_b{ T^j})} \right)\,,\nonumber\eeqa  with
\beqa
V({T^iT^i})&=&e^{-\pi{ T^iT^i}/2}, \quad Q^{ij}=I\delta^{ij}-i[T^i,T^j],\eeqa

where $i,j=1,2$, \ie $T^1=T\sigma_1$, $T^2=T\sigma_2$ . 
The entire information about this effective action is given in  \cite{Hatefi:2013yxa}.  This action \reef{nonab} is also consistent with the entire S-matrix calculations of $<V_C V_TV_{\phi}V_{\phi}>$ of \cite{Hatefi:2012wj}.

Around the stable point of  tachyon potential this action gets reduced to tachyon DBI action \cite{Sen:1999md} with $T^4V(T^2)$ potential and by taking the limit of the tachyon  to infinity, the term  $T^4V(TT)$ is sent  to zero and this can be well described from  condensation of  an unstable brane.
\vskip.1in

By making use of the S-matrix of $<V_C V_{\phi}V_TV_T>$ in \cite{Hatefi:2012cp} and $<V_C V_{A}V_TV_T>$  in \cite{Garousi:2007fk},
we have also explored not only the entire  form of  D-brane -anti D-brane effective actions but also its all order $\alpha'$ corrections of two scalar (two gauge field)-two tachyon couplings.

Various applications in the literature in favour of those higher derivative string couplings  have been pointed out. For instance, employing either some  new BPS (non-BPS) couplings, their corrections or Myers terms to M-theory \cite{Horava:1995qa} one can actually  interpret and get to  the phenomenon of $N^3$ entropy growth of $M5$ branes or discuss various combinations of $M2,M5$ branes\cite{Hatefi:2012sy}. Evaluating some of (non)-BPS couplings , one derives (AdS)-dS brane world solutions \cite{Hatefi:2012bp} and could further elaborate on the point that  despite the fact that we are dealing with   non-supersymmetric case, as long as the EFT holds, all the large volume scenario minima will become stable \cite{deAlwis:2013gka}.  Applying CFT techniques \cite{Kennedy:1999nn} the tachyonic DBI supersymmetrized action was suggested \cite{Aganagic:1997zk} to be as follows
\beqa
L=-T_pV(T)\sqrt{-\det(\eta_{ab}+2\pi\alpha'F_{ab}-2\pi\alpha'\bar\Psi\ga_b\prt_a\Psi+\pi^2\alpha'^2\bar\Psi\ga^{\mu}\prt_a\Psi\bar\Psi\ga_{\mu}\prt_b\Psi + 2\pi\alpha'\prt_aT\prt_bT )}\label{esi111a}\eeqa
%\nonumber
%\label{esi111}
%\eeqa

 One of the reasons we work with non-BPS branes is due to its direct relationship with realising  the properties IIA(IIB) 
 string theory in almost time-dependent backgrounds \cite{Sen:2002in}.

 The source of instability in flat empty space is tight to the presence of tachyons and by using their
 effective action  one hopes to get diverse results , such as the evaluation of these non-BPS branes and indeed tachyonic 
 action can well describe decay properties of unstable D$_p$-branes. 
   %\cite{Sen:2002in,Sen:2002an}. 
 
 The cosmological applications through these effective actions can also be addressed , 
 for example if one considers, the action of D-brane anti D-brane then one is able to explain
 inflation through string theory \cite{Kachru:2003sx} whereas several other applications to non-BPS 
 branes have also been released \cite{Frau:1999qs}. Tachyons and their corrections can also be
 employed for models such as holographic like QCD \cite{Casero:2007ae} and finally brane anti brane system as a background was used in Sakai-Sugimoto models \cite{Sakai:2004cn}.
 %If we take the brane, anti brane system just as a background (to be dual to confined color theory) then one can introduce flavor branes \cite{Sakai:2004cn} where by decreasing the number of  flavor branes with respect to the number of color branes, the $D\bar D$  system may be  considered as a probe. 
 \vskip .1in

The outline of  the paper is as follows. 
 \vskip .1in
 
In the next section ,  we first introduce all the vertex operators with their CP matrix,  apply CFT techniques to actually obtain the entire form of a symmetric RR, a scalar, a gauge and  a tachyon of $<V_{C^{-1}} V_{\phi^{0}}V_{A^{0}}V_{T^{-1}}>$ S-matrix  in both transverse and world volume directions of non-BPS branes of IIA(IIB) superstring theory.  We then make use of the  expansion for non-BPS branes and reveal that apart from an infinite $u'=u+\frac{1}{4}$ and ($t$ ) channel  tachyon (scalar) singularities, this S-matrix clearly does involve an infinite $u',(t)$ channel  singularity structures  in the bulk  as well. Indeed we show that
${\cal A}_{3},{\cal A}_{4}$ of this S-matrix do carry an infinite $p.\xi_1$ singular terms , whose momenta are located in the bulk directions and from now on due to the presence of $p^i$, we are going to call them bulk singularities and start producing them in effective field theory (EFT) as well.

We also write down the results for symmetric $<V_{C^{-1}} V_{\phi^{-1}}V_{A^{0}}V_{T^{0}}>$ ,$<V_{C^{-1}} V_{\phi^{0}}V_{A^{-1}}V_{T^{0}}>$ and asymmetric $<V_{C^{-2}} V_{\phi^{0}}V_{A^{0}}V_{T^{0}}>$ S-matrix  and start comparing them which leads to finding out various generalised Bianchi Identities in the presence of non-BPS branes. In section 2 and 3 we produce an infinite number of $u'$, $t$ channel tachyon , scalar singularities.  At the end we apply all order $\alpha'$ higher derivative corrections properly to Chern-Simons couplings in such a way that even we are able to produce an infinite number of  of $u'$, $t$ channel bulk singularity structures. This obviously confirms that those bulk singularities are also needed in the entire S-matrix as they even can be constructed in EFT side as well.

Note that based on various results of this paper, we reveal the important fact as follows.  A-priori  without knowing any  restricted Bianchi identities on world volume of D-branes for RR , there seems to be no chance to even see  all the needed  bulk singularity structures of ${\cal A}_{3}$ and  ${\cal A}_{4}$ of \reef{48} from  the other symmetric analysis (unlike asymmetric analysis) as we go through them in detail in the next sections.

\subsection{ The entire  $<C^{-1}\phi^{0}A^{0}T^{-1}>$ S-matrix }

Here we would like to explore the entire form of the S-matrix elements (both in transverse and world volume direction) of an RR, a real massless scalar field, a gauge field and a real tachyon of type IIA (IIB) superstring theory, which is indeed a five point non-BPS S-matrix from the world sheet point of view. First we  are going to explain our notations so that

all $\mu,\nu,...$ show the entire ten dimensional space-time, on the other hand all  $a, b, c, .. $and $i, j, k, ..$ are taken to be employed for world volume and transverse directions appropriately.

It is discussed in \cite{Hatefi:2012wj} that in the presence of non-BPS branes one needs to consider the Chan-Paton matrices inside the vertices as follows 
 \beqa
V_{T}^{(0)}(x) &=&  \alpha' ik\cd\psi(x) e^{\alpha' ik\cd X(x)}\lam\otimes\sigma_1,
\nonumber\\
V_{T}^{(-1)}(x) &=&e^{-\phi(x)} e^{\alpha' ik\cd X(x)}\lam\otimes\sigma_2\nonumber\\
V_\phi^{(-1)}(x)&=&e^{-\phi(x)}\xi_i\psi^i(x)e^{ \alpha'iq\inn X(x)}\lam\otimes \sigma_3 \nonumber\\
V_A^{(-1)}(x)&=&e^{-\phi(x)}\xi_a\psi^a(x)e^{ \alpha'iq\inn X(x)}\lam\otimes \sigma_3 \nonumber\\
V_{\phi}^{(0)}(x) &=& \xi_{1i}(\partial X^i(x)+i\alpha'k.\psi\psi^i(x))e^{\alpha'ik.X(x)}\otimes I\nonumber\\
V_{\phi}^{(-2)}(x) &=& e^{-2\phi(x)}V_{\phi}^{(0)}(x)\otimes I\nonumber\\
V_{A}^{(0)}(x) &=& \xi_{1a}(\partial X^a(x)+i\alpha'k.\psi\psi^a(x))e^{\alpha'ik.X(x)}\otimes I\label{d4Vs}\\
%V_{\bar\Psi}^{(-1/2)}(x)&=&\bar u^Ae^{-\phi(x)/2}S_A(x)\,e^{ \alpha'iq.X(x)}\lam\otimes\sigma_3 \nonumber\\
%V_{\Psi}^{(-1/2)}(x)&=&u^Be^{-\phi(x)/2}S_B(x)\,e^{ \alpha'  iq.X(x)}\lam\otimes I \nonumber\\
V_{C}^{(-\frac{1}{2},-\frac{1}{2})}(z,\bar{z})&=&(P_{-}\fsH_{(n)}M_p)^{\alpha\beta}e^{-\phi(z)/2}
S_{\al}(z)e^{i\frac{\alpha'}{2}p\cd X(z)}e^{-\phi(\bar{z})/2} S_{\be}(\bar{z})
e^{i\frac{\alpha'}{2}p\cd D \cd X(\bar{z})}\lam\otimes\sigma_3\sigma_1,\nonumber\\
V_{C}^{(-\frac{3}{2},-\frac{1}{2})}(z,\bar{z})&=&(P_{-}\slashed{C}_{(n-1)}M_p)^{\alpha\beta}e^{-3\phi(z)/2}
S_{\al}(z)e^{i\frac{\alpha'}{2}p\cd X(z)}e^{-\phi(\bar{z})/2} S_{\be}(\bar{z})
e^{i\frac{\alpha'}{2}p\cd D \cd X(\bar{z})}\lam\otimes\sigma_1\nonumber\eeqa

 Note that the CP factors of RR   in the presence of $D\bar D$ system  for symmetric and asymmetric pictures  are $\sigma_3$ and $I$ accordingly. Hence   $<C^{-1}\phi^{0}A^{0}T^{-1}>$ S-matrix  shall be taken as follows

\beqa
{\cal A}^{<C^{-1}\phi^{0}A^{0}T^{-1}>} & \sim & \int dx_{1}dx_{2}dx_{3}dzd\bar{z}\,
  \lan V_{\phi}^{(0)}{(x_{1})}
V_{A}^{(0)}{(x_{2})}V_T^{(-1)}{(x_{3})}
V_{RR}^{(-\frac{1}{2},-\frac{1}{2})}(z,\bar{z})\ran,\labell{cor10}\eeqa

where we are dealing with the disk level amplitude and mass-shell conditions for $k_1,k_2,p,k_3$ are
\beqa
 k_{1}^2=k_{2}^2=p^2=0, \quad  k_{3}^2=\frac{1}{2\alpha'}  ,\quad  k_2.\xi_2=k_2.\xi_1=k_1.\xi_1=k_3.\xi_1=0
\label{8800aa}\eeqa
where we set $\alpha'=2$ .
In order to just use the holomorphic correlators, one needs to keep track of the following notations for projection, RR field strength and for spinor as well.

\begin{displaymath}
P_{-} =\ha (1-\ga^{11}), \quad
\fsH_{(n)} = \frac{a
_n}{n!}H_{\mu_{1}\ldots\mu_{n}}\ga^{\mu_{1}}\ldots
\ga^{\mu_{n}},\quad
(P_{-}\fsH_{(n)})^{\al\be} =
C^{\al\del}(P_{-}\fsH_{(n)})_{\del}{}^{\be}.
\non\end{displaymath}

For type IIA  (type IIB) $n=2,4$,$a_n=i$  ($n=1,3,5$,$a_n=1$) .

We also apply the doubling trick to make use of  just holomorphic parts of all the world sheet fields as below  
\begin{displaymath}
\tilde{X}^{\mu}(\bar{z}) \rightarrow D^{\mu}_{\nu}X^{\nu}(\bar{z}) \ ,
\spa
\tilde{\psi}^{\mu}(\bar{z}) \rightarrow
D^{\mu}_{\nu}\psi^{\nu}(\bar{z}) \ ,
\spa
\tilde{\phi}(\bar{z}) \rightarrow \phi(\bar{z})\,, \mand
\tilde{S}_{\al}(\bar{z}) \rightarrow M_{\al}{}^{\be}{S}_{\be}(\bar{z})
 \ ,
\non\end{displaymath}

with the following definitions for the aforementioned matrices
\begin{displaymath}
D = \left( \begin{array}{cc}
-1_{9-p} & 0 \\
0 & 1_{p+1}
\end{array}
\right) \ ,\,\, \mand
M_p = \left\{\begin{array}{cc}\frac{\pm i}{(p+1)!}\ga^{i_{1}}\ga^{i_{2}}\ldots \ga^{i_{p+1}}
\eps_{i_{1}\ldots i_{p+1}}\,\,\,\,{\rm for\, p \,even}\\ \frac{\pm 1}{(p+1)!}\ga^{i_{1}}\ga^{i_{2}}\ldots \ga^{i_{p+1}}\ga_{11}
\eps_{i_{1}\ldots i_{p+1}} \,\,\,\,{\rm for\, p \,odd}\end{array}\right.
\non\end{displaymath}
\vskip .2in
Having set that, we can now go ahead with the correct form of the correlations for all  $X^{\mu},\psi^\mu, \phi$ fields , as follows
\begin{eqnarray}
\lan X^{\mu}(z)X^{\nu}(w)\ran & = & -\frac{\alpha'}{2}\eta^{\mu\nu}\log(z-w) \ , \non \\
\lan \psi^{\mu}(z)\psi^{\nu}(w) \ran & = & -\frac{\alpha'}{2}\eta^{\mu\nu}(z-w)^{-1} \ ,\non \\
%\lan c(z)c(w) \ran & = & (z-w) \ , \non \\
\lan\phi(z)\phi(w)\ran & = & -\log(z-w) \ .
\labell{prop2}\end{eqnarray}

Substituting the above vertex operators into the S-matrix, the amplitude reads off 
\beqa {\cal A}^{<C^{-1}\phi^{0}A^{0}T^{-1}>}&\sim& \int
 dx_{1}dx_{2}dx_{3}dx_{4} dx_{5}\,
(P_{-}\fsH_{(n)}M_p)^{\al\be}\xi_{1i}\xi_{2a}x_{45}^{-1/4}(x_{34}x_{35})^{-1/2}\nonumber\\&&
\times(I_1+I_2+I_3+I_4)\Tr(\lam_1\lam_2\lam_3)\Tr(\sig_3\sig_1\sig_2),\labell{1255}\eeqa
so that one needs to go over the following correlation functions 
\beqa
I_1&=&{<:\partial X^i{(x_1)}e^{\alpha'ik_1.X(x_1)}: \partial X^a{(x_2)}e^{\alpha'ik_2.X(x_2)}
 :e^{\alpha'ik_3.X(x_3)}:e^{\frac{\alpha'}{2}ip.X(x_4)}:e^{\frac{\alpha'}{2}ip.D.X(x_5)}:>}\nonumber \\&&\times{<:S_{\al}(x_4):S_{\be}(x_5):>},\nonumber\\
I_2&=&{<:\partial X^i{(x_1)}e^{\alpha'ik_1.X(x_1)}: e^{\alpha'ik_2.X(x_2)}
 :e^{\alpha'ik_3.X(x_3)}:e^{\frac{\alpha'}{2}ip.X(x_4)}:e^{\frac{\alpha'}{2}ip.D.X(x_5)}:>}\nonumber \\&&
\alpha'ik_{2b} {<:S_{\al}(x_4):S_{\be}(x_5):\psi^b\psi^{a}(x_2):>}\nonumber \\
I_3&=&{<:e^{\alpha'ik_1.X(x_1)}: \partial X^a{(x_2)}e^{\alpha'ik_2.X(x_2)}
 :e^{\alpha'ik_3.X(x_3)}:e^{\frac{\alpha'}{2}ip.X(x_4)}:e^{\frac{\alpha'}{2}ip.D.X(x_5)}:>}\nonumber \\&&
\times\alpha'ik_{1c} {<:S_{\al}(x_4):S_{\be}(x_5):\psi^c\psi^{i}(x_1):>}\nonumber \\
I_4&=&{<:e^{\alpha'ik_1.X(x_1)}: e^{\alpha'ik_2.X(x_2)}
 :e^{\alpha'ik_3.X(x_3)}:e^{\frac{\alpha'}{2}ip.X(x_4)}:e^{\frac{\alpha'}{2}ip.D.X(x_5)}:>}\nonumber \\&&
\times(-(\alpha')^2k_{1c}k_{2b}) {<:S_{\al}(x_4):S_{\be}(x_5):\psi^c\psi^{i}(x_1):\psi^b\psi^{a}(x_2):>}
\eeqa

Having taken the  Wick-theorem and \reef{prop2}, we were able to compute all the  correlators of $X$.  However, to get to fermionic correlations , involving the spin operators , one employs the  so called Wick-like rule~\cite{Liu:2001qa}  as follows

\beqa
I_2^{ic}&=&<:S_{\al}(x_4):S_{\be}(x_5):\psi^c\psi^i(x_1):>
=2^{-1}x_{45}^{-1/4} (x_{14}x_{15})^{-1}(\Gamma^{ic}C^{-1})_{\alpha\beta}
\nonumber\\
%\label{68}\eeqa\beqa
I_3^{ab}&=&<:S_{\al}(x_4):S_{\be}(x_5):\psi^b\psi^a(x_2):>
=2^{-1}x_{45}^{-1/4} (x_{24}x_{25})^{-1}(\Gamma^{ab}C^{-1})_{\alpha\beta}.
\label{691}\eeqa
where $x_{ij}=x_i-x_j, x_4=z,x_5=\bar z$. For the calculations of two spin operators and two currents one applies  the generalization of Wick-Like rule ~\cite{Hatefi:2015jpa}  to indeed find out the fermionic correlations as follows

\beqa
I_4^{abic}&=&<:S_{\al}(x_4):S_{\be}(x_5):\psi^c\psi^{i}(x_1):\psi^b\psi^{a}(x_2):>\nonumber\\&&
=\bigg\{(\Gamma^{abic}C^{-1})_{{\alpha\beta}}+\alpha'\frac{Re[x_{14}x_{25}]}{x_{12}x_{45}}(\eta^{cb}(\Gamma^{ai}C^{-1})_{{\alpha\beta}}-\eta^{ac}(\Gamma^{bi}C^{-1})_{{\alpha\beta}})\bigg\}\nonumber\\&&
2^{-2}x_{45}^{3/4}(x_{14}x_{15}x_{24}x_{25})^{-1}\label{hh}\eeqa

Considering all the bosonic and fermionic correlators into  \reef{1255} , one reveals the whole closed part of the amplitude as follows  
\beqa
{\cal A}^{<C^{-1}\phi^{0} A^{0}T^{-1}>}&\!\!\!\!\sim\!\!\!\!\!&\int dx_{1}dx_{2} dx_{3}dx_{4}dx_{5}(P_{-}\fsH_{(n)}M_p)^{\al\be}I\xi_{1i}\xi_{2a}(2i) x_{45}^{-1/4}(x_{34}x_{35})^{-1/2}\nonumber\\&&\times
\bigg(a^i_1a^a_2 x_{45}^{-5/4} C^{-1}_{\alpha\beta}+\alpha'ik_{2b}a^i_1I_3^{ab}+\alpha'ik_{1c}a^a_2I_2^{ic}-(\alpha')^2k_{1c}k_{2b}I_4^{abic}\bigg)\Tr(\lam_1\lam_2\lam_3)\nonumber\eeqa
%where  $I_4^{cadi}$ is given in \reef{hh} and
such that
\beqa
I&=&|x_{12}|^{\alpha'^2k_1.k_2}|x_{13}|^{\alpha'^2k_1.k_3}|x_{14}x_{15}|^{\frac{\alpha'^2}{2}k_1.p}|x_{23}|^{\alpha'^2k_2.k_3}|x_{24}x_{25}|^{ \frac{\alpha'^2}{2}  k_2.p}
|x_{34}x_{35}|^{\frac{\alpha'^2}{2}   k_3.p}|x_{45}|^{\frac{\alpha'^2}{4}    p.D.p},\nonumber\\
a^i_1&=&ip^i\frac{x_{54}}{x_{14}x_{15}},\nonumber\\
a^a_2&=&ik_1^{a}\bigg(\frac{x_{14}}{x_{12}x_{24}}+\frac{x_{15}}{x_{12}x_{25}}\bigg)
+ik_3^{a}\bigg(\frac{x_{43}}{x_{23}x_{24}}+\frac{x_{53}}{x_{23}x_{25}}\bigg).
\eeqa

One is able now to precisely check out the  SL(2,R) invariance of the above S-matrix, and we do  gauge fixing  by fixing  all the positions of open strings as 
 $x_{1}=0, x_{2}=1, x_{3}\rightarrow \infty$ %\qquad dx_1dx_2dx_3\rightarrow x_3^{2},
 so that at the end, one has to come over the following sort of integration on the upper half complex plane 
\beqa
 \int d^2 \!z |1-z|^{a} |z|^{b} (z - \bar{z})^{c}
(z + \bar{z})^{d},
 \eeqa
Note that all   $a,b,c$ are some combinatoric parts of the defined Mandelstam variables as follows
\beqar
s&=&-(k_1+k_3)^2,\qquad t=-(k_1+k_2)^2,\qquad u=-(k_2+k_3)^2.
\label{eerr22}\eeqar
Results of integrations for both  $d=0,1$ and for $d=2$ have been appropriately explored  in  \cite{Fotopoulos:2001pt}, \cite{Hatefi:2012wj}, so that the compact and ultimate results for the entire S-matrix in both transverse and world volume directions of brane are discovered as 
\beqa {\cal A}^{<C^{-1}\phi^{0} A^{0}T^{-1}>}&=&{\cal A}_{1}+{\cal A}_{2}+{\cal A}_{3}+{\cal A}_{4}+{\cal A}_{5}+{\cal A}_{6},\labell{15u}\eeqa
where
\beqa
{\cal A}_{1}&\!\!\!\sim\!\!\!&2i\bigg(-\xi_{1i}\xi_{2a}k_{1c}k_{2b}
\Tr(P_{-}\fsH_{(n)}M_p\Gamma^{abic})+\xi_1.p\xi_{2a} k_{2b}\Tr(P_{-}\fsH_{(n)}M_p\Gamma^{ab})
\bigg)L_1,
\nonumber\\
{\cal A}_{2}&\!\!\!\sim\!\!\!&2i\xi_{1i}\xi_{2a}
\Tr(P_{-}\fsH_{(n)}M_p\Gamma^{ai}) (t)(u+\frac{1}{4})L_3,
\nonumber\\
{\cal A}_{3}&\!\!\!\sim\!\!\!&4ik_1.\xi_2 \xi_1.p\Tr(P_{-}\fsH_{(n)}M_p) (u+\frac{1}{4})L_3,
\nonumber\\
{\cal A}_{4}&\!\!\!\sim\!\!\!&-4ik_3.\xi_2 \xi_1.p\Tr(P_{-}\fsH_{(n)}M_p) (t)L_3,
\nonumber\\
{\cal A}_{5}&\!\!\!\sim\!\!\!&4ik_3.\xi_2 \xi_{1i}k_{1c}\Tr(P_{-}\fsH_{(n)}M_p\Gamma^{ic}) (t)L_3,
\nonumber\\
{\cal A}_{6}&\!\!\!\sim\!\!\!&4ik_1.\xi_2 (k_{1c}+k_{2c})\xi_{1i}\Tr(P_{-}\fsH_{(n)}M_p\Gamma^{ic}) (u+\frac{1}{4})L_3,
\labell{48}\eeqa

where

\beqa
L_1&=&(2)^{-2(t+s+u)}\pi{\frac{\Gamma(-u+\frac{1}{4})
\Gamma(-s+\frac{1}{4})\Gamma(-t+\frac{1}{2})\Gamma(-t-s-u+\frac{1}{2})}
{\Gamma(-u-t+\frac{3}{4})\Gamma(-t-s+\frac{3}{4})\Gamma(-s-u+\frac{1}{2})}},\nonumber\\
L_3&=&(2)^{-2(t+s+u)-1}\pi{\frac{\Gamma(-u-\frac{1}{4})
\Gamma(-s+\frac{3}{4})\Gamma(-t)\Gamma(-t-s-u)}
{\Gamma(-u-t+\frac{3}{4})\Gamma(-t-s+\frac{3}{4})\Gamma(-s-u+\frac{1}{2})}}.
\nonumber\eeqa

This S-matrix does satisfy  the Ward identity associated to the gauge field, so that by replacing   $\xi_{2a}\rightarrow k_{2a}$  the whole S-matrix vanishes.
One can also use $(k_1+k_2)_{c}=(k_3+p)_{c}$in ${\cal A}_{6}$ .

Note that , if we just change the picture of scalar field in the presence of a symmetric RR,  one ends up having the final form of the S-matrix of $<C^{-1}\phi^{-1} A^{0}T^{0}>$ \cite{Hatefi:2013yxa} as follows

\beqa {\cal A}^{<C^{-1}\phi^{-1} A^{0}T^{0}>}&=&{\cal A}_{1}+{\cal A}_{2}+{\cal A}_{3},\labell{17u}\eeqa
where
\beqa
{\cal A}_{1}&\!\!\!\sim\!\!\!&2\xi_{1i}\xi_{2a}k_{3c}k_{2d}
\Tr(P_{-}\fsH_{(n)}M_p\Gamma^{cadi}
)L_1,
\nonumber\\
{\cal A}_{2}&\sim& \bigg\{-\Tr(P_{-}\fsH_{(n)}M_p \gamma.\xi_2\gamma.\xi_{1})(u+\frac{1}{4})-2k_3.\xi_2\Tr(P_{-}\fsH_{(n)}M_p \gamma.k_2\gamma.\xi_{1})\bigg\}L_3(2t)
\nonumber\\&&
+\Tr(P_{-}\fsH_{(n)}M_p \gamma.k_3\gamma.\xi_{1})\bigg\{2t(k_3.\xi_2)+2(-u-\frac{1}{4})k_1.\xi_2\bigg\}(-2L_3).
\labell{487}\eeqa

  \vskip.2in
  If one applies the momentum conservation along the world volume of brane, one then realises the fact that ${\cal A}_{1}$ of \reef{487}
produces the first term  ${\cal A}_{1}$ of \reef{48}. The first term  ${\cal A}_{2}$ of \reef{487}
exactly generates  ${\cal A}_{2}$ of \reef{48}, the sum of the second and third term  ${\cal A}_{2}$ of \reef{487}, reconstructs ${\cal A}_{5}$ of \reef{48} and finally the last term  ${\cal A}_{2}$ of \reef{487} generates ${\cal A}_{6}$ of \reef{48}. Thus there seems to be no chance to produce all the needed  bulk singularities  ${\cal A}_{3}$ and  ${\cal A}_{4}$ of \reef{48}. It is also important to stress that the second term   ${\cal A}_{1}$ of \reef{48} was also overlooked in \reef{487}.
  \vskip.1in
   
   Note that since the momentum along the brane  is just conserved, from now on instead of Bianchi identities, we use  Bianchi identities restricted on D-brane directions or  restricted Bianchi identities on world volume. Consider the following restricted  Bianchi identity on world volume as follows \beqa
  p^i \eps^{a_{0}\cdots a_{p}}H_{a_{0}\cdots a_{p}}-p_c\eps^{a_{0}\cdots a_{p-1}c}H^{i}_{a_{0}\cdots a_{p-1}}&=&0
 \label{esi744}
 \eeqa
 
 Note that $H$  in \reef{esi744}  is  $(p+1)$ form field strength of $C_p$  form and this is obviously true from the traces of gamma matrices which appear in all ${\cal A}_{2}$  terms of \reef{487}, basically all the traces for ${\cal A}_{2}$  terms of \reef{487}  are non-zero just for  $n=p+1$ case.
 
 %Hp?1 of a Cp?2 form and the field strength appearing in . So the terms in his various equations (17) etc correspond to different RR field couplings. 
 
 Upon taking into account the above restricted  Bianchi identity and applying momentum conservation along the world volume of brane $(k_1+k_2+k_3+p)^a=0$ to the sum of the 2nd and 3rd term of ${\cal A}_{2}$  (and also to the last term of ${\cal A}_{2}$) of \reef{487}, one is able to  actually produce precisely  all infinite $u'$ ($t$) channel bulk singularities ${\cal A}_{4}$  (and ${\cal A}_{3}$)  of \reef{48} accordingly.
\vskip.1in

While a priori  without knowing any restricted Bianchi identity on world volume for RR , there seems to be no chance to even see  all the needed  bulk singularities of ${\cal A}_{3}$ and  ${\cal A}_{4}$ of \reef{48} from  $<C^{-1}\phi^{-1}A^{0}T^{0}>$ S-matrix.

\vskip.1in

Let us see what happens in the other picture of the S-matrix. One reads off the S-matrix $<C^{-1}\phi^{0}A^{-1}T^{0}>$  \cite{Hatefi:2015gwa} as follows

\beqa {\cal A}^{<C^{-1}\phi^{0}A^{-1}T^{0}>}&=&{\cal A}_{1}+{\cal A}_{2},\labell{11u}\eeqa

with
\beqa
{\cal A}_{1}&\!\!\!\sim\!\!\!&\bigg(2\xi_{1i}\xi_{2a}k_{3c}k_{1d}
\Tr(P_{-}\fsH_{(n)}M_p\Gamma^{caid})-\xi_1.p(2k_{3c}\xi_{2a}) \Tr(P_{-}\fsH_{(n)}M_p\Gamma^{ca})
\bigg)L_1,
\nonumber\\
{\cal A}_{2}&\sim&  \bigg\{t\xi_1.p (4k_{3}.\xi_2)\Tr(P_{-}\fsH_{(n)}M_p )+ 4(u+\frac{1}{4})k_{3c}\xi_{1i}\Tr(P_{-}\fsH_{(n)}M_p\Gamma^{ci})k_1.\xi_2\nonumber\\&&-4tk_{3}.\xi_{2} k_{1b}\xi_{1i}\Tr(P_{-}\fsH_{(n)}M_p\Gamma^{ib})-2t(u+\frac{1}{4}) \xi_{1i}\xi_{2a}\Tr(P_{-}\fsH_{(n)}M_p\Gamma^{ai})
\bigg\}L_3
\labell{497m}\eeqa
 
  \vskip.2in

By comparisons of the elements of \reef{497m} with   \reef{48}, we are able to produce all the terms inside \reef{48} except its ${\cal A}_{3}$. In the other words again in this $<C^{-1}\phi^{0}A^{-1}T^{0}>$ S-matrix , there seems to be no chance to produce ${\cal A}_{3}$   bulk singularities of \reef{48}.

Taking into account the above  restricted Bianchi identity \reef{esi744} and applying momentum conservation along the world volume of brane  to the  2nd term ${\cal A}_{2}$  of \reef{497m}, one is able to  indeed construct exactly  all infinite  $t$ channel bulk singularities ${\cal A}_{3}$  of \reef{48}. Meanwhile in this particular picture of S-matrix ($<C^{-1}\phi^{0}A^{-1}T^{0}>$)   one could already see that  the infinite $u'$ channel bulk singularities   ${\cal A}_{4}$ of  \reef{48} have been shown up in the entire S-matrix.

While a priori  without knowing any  restricted Bianchi identity for RR , there seems to be no chance to even observe all the needed  
$t$ channel bulk singularities ${\cal A}_{3}$  of \reef{48} from  $<C^{-1}\phi^{0}A^{-1}T^{0}>$ S-matrix.
 \vskip.06in

Note that if we would consider the Ward identity associated to the gauge field $(\xi_{2a}\rightarrow k_{2a})$, we would reveal that due to presence of the 2nd term of ${\cal A}_{1}$ of \reef{497m} and 1st term  ${\cal A}_{2}$ of \reef{497m} , the S-matrix is  not  gauge invariant any more. In order to restore gauge invariance, one needs to consider further remarks. Basically if we replace  $(\xi_{2a}\rightarrow k_{2a})$ in all four terms   ${\cal A}_{2}$ of \reef{497m}, apply momentum conservation along the world volume of brane as well as  simultaneously consider the restricted Bianchi identity \reef{esi744}, then one observes that all four terms ${\cal A}_{2}$ of \reef{497m} respect Ward identity.

Finally if one replaces  $(\xi_{2a}\rightarrow k_{2a})$ in  all two terms   ${\cal A}_{1}$ of \reef{497m} , apply momentum conservation along the world volume of brane as well as  take into account the following restricted Bianchi identity on world volume directions 
\beqa
\xi_{1i} k_{3c} k_{2a}(-p_d\eps^{a_{0}\cdots a_{p-3}cad}H^{i}_{a_{0}\cdots a_{p-3}}+p^i \eps^{a_{0}\cdots a_{p-2}ac}H_{a_{0}\cdots a_{p-2}})&=&0 \label{ccbbdd}\eeqa
then one obviously clarifies  that all two terms   ${\cal A}_{1}$ of \reef{497m} are also now respecting Ward identity associated to gauge field, therefore based on applying those restricted Bianchi identities on world volume, now the whole S-matrix respects Ward identity.

 \vskip.1in
 
 Note that the form that appears in \reef{ccbbdd}  is  $H$ of $(p-1)$ form field strength of $C_{p-2}$  form and this is  true from the traces of gamma matrices which appear in all ${\cal A}_{1}$  terms of \reef{497m}, basically all the traces for ${\cal A}_{1}$  terms of \reef{497m}  are non-zero just for  $n+1=p$ case. Hence, the terms in  \reef{esi744} and  \reef{ccbbdd} do correspond to different RR field couplings.

It is worth mentioning that unlike the symmetric picture , in  the asymmetric picture of  RR and in the presence of non-BPS branes, without using any further restricted Bianchi identity the ultimate result of amplitude does satisfy  Ward identity associated to the gauge field.
Eventually one can compute the same S-matrix but in  asymmetric picture of RR so that the result of $<C^{-2}\phi^{0}A^{0}T^{0}>$ is found in \cite{Hatefi:2015gwa}  to be
 \beqa {\cal A}^{<C^{-2}\phi^{0}A^{0}T^{0}>}&=&{\cal A}_{1}+{\cal A}_{2}+{\cal A}_{3}+{\cal A}_{4}\labell{1u8u}\eeqa
where
\beqa
{\cal A}_{1}&\sim & 2^{3/2}i\xi_{1i}\xi_{2a}k_{3c}k_{2b}L_1
\bigg(p^i\Tr(P_{-}\slashed C_{(n-1)}M_p\Gamma^{cab})-k_{1d}\Tr(P_{-}\slashed C_{(n-1)}M_p\Gamma^{cabid})\bigg)
\nonumber\\
{\cal A}_{2}&\sim&2^{3/2}i\xi_1.p L_3 \Tr(P_{-}\slashed C_{(n-1)}M_p \gamma^c)\bigg(2tk_3.\xi_2  [-k_{3c}-k_{2c}]
+2k_1.\xi_2u'k_{3c}-tu'\xi_{2c}\bigg)
\nonumber\\
{\cal A}_{3}&\sim &2^{3/2}i\xi_{1i}L_3 \Tr(P_{-} \slashed C_{(n-1)}M_p \Gamma^{cid})\bigg[-2k_1.\xi_2 u'k_{3c}(k_{1d}+k_{2d})+2tk_3.\xi_2 k_{1d}(k_{3c}+k_{2c})\bigg]
\nonumber\\
{\cal A}_{4}&\sim &2^{3/2}i\xi_{1i}L_3t u' \xi_{2a} \Tr(P_{-}  \slashed C_{(n-1)}M_p \Gamma^{cai})(k_{3c}+k_{1c}+k_{2c})
\labell{qq689n}\eeqa

  As we can see in this asymmetric picture we can precisely produce even all the bulk singularities ${\cal A}_{3}$ and 
  ${\cal A}_{4}$ of \reef{48}.

\vskip.1in
Considering all the definitions of $\fsH_{(n)}, M_p, \Gamma^{cadi}$ , one knows that the S-matrix is non zero just for  $p=n+1$ and $ p+1=n$ cases.
Keeping in mind the momentum conservation along the world volume of brane, one obtains 
\beqa
s+t+u=-p^ap_a-\frac{1}{4},\label{momss}\eeqa

Standard scattering of RR on BPS branes, with three massless open string states, does not restrict  $p^ap_a$ momentum invariant of the closed string state to a specific value. Obviously for non-BPS branes due to the presence of the tachyon the kinematic invariants  are more restricted. Indeed for an RR and a tachyon momentum conservation leads to  $p^ap_a=k^2=\frac{1}{4}$.  It is also discussed in \cite{Hatefi:2012wj} that for an RR, two massless open strings and  a tachyon, using the on-shell relations $k_1^{2}=k_2^{2}=0$ and $k_3^{2}=\frac{1}{4}$, we are able to rewrite  the momentum expansion
\beqa
  (k_1+k_2)^2 \rightarrow 0,  \quad  k_1.k_3 \rightarrow  0 , \quad k_2.k_3 \rightarrow  0,
\label{momss22}\eeqa 
just in terms of $t \rightarrow 0,  \quad  s \rightarrow  -\frac{1}{4} , \quad u \rightarrow  -\frac{1}{4}$. 
In fact  the constraint \reef{momss} clearly confirms that $p^ap_a $ must be sent to $ \frac{1}{4}$ or $p^ap_a \rightarrow \frac{1}{4}$ and this  just  makes sense only for euclidean brane. This is also consistent with the observation  that has been pointed out  in \cite{Billo:1999tv}, which means that  on-shell conditions do impose to us the fact that the amplitude must be carried out just for  non-BPS SD-branes \cite{Gutperle:2002ai22}. The other point which is worth mentioning is as follows. It is shown in \cite{Hatefi:2012wj} that  the constraint  $p^ap_a \rightarrow \frac{1}{4}$ is valid for all three, four and five point non-BPS functions and more importantly it is checked that by using the constraint$p^ap_a =\frac{1}{4}$, one is able to precisely produce all infinite  $(t+s+u+\frac{1}{2})$ tachyon poles of  an RR, two scalar fields and a tachyon of  \cite{Hatefi:2012wj} as the final form of amplitude clearly involves the factor $\Gamma(-t-s-u-\frac{1}{2})$ ( for more information see equation (20) and section 4 of \cite{Hatefi:2012wj}).
\vskip.1in
Now using  $p^ap_a \rightarrow \frac{1}{4}$ for non-BPS branes , also taking into account the non zero   vertex operator of two scalars and a  gauge field , one immediately gets to know that  $t \rightarrow 0,  \quad  s \rightarrow  -\frac{1}{4} , \quad u \rightarrow  -\frac{1}{4}$ is the only unique expansion of the S-matrix.  Given the facts that the S-matrix does not include the coefficients of $\Gamma(-s-\frac{1}{4}), \Gamma(-t-s-u-\frac{1}{2})$, and also $ <V_{A^{0}} V_{T^{0}} V_{\phi^{-1}} V_{\phi^{-1}}>, <V_{A^{0}} V_{T^{0}} V_{A^{-1}} V_{\phi^{-1}}>$ have zero contribution (based on applying CP matrices , as  $\Tr(I\sig_1\sig_3\sig_3)=0$)
, we understand that, this S-matrix does not have any $s'=s+1/4, (s'+t+u')$ poles at all, thus we are left over with an infinite number of $u',t$ channel poles.

\vskip.1in
Let us analyze all infinite tachyon $u'$ channel poles , then we reconstruct all infinite t-channel scalar poles accordingly.
Indeed, by taking the momentum expansion into considerations, we realise that  ${\cal A}_{1},{\cal A}_{2}$ in \reef{15u} are all contact interactions, while ${\cal A}_{3},{\cal A}_{6}$  and ${\cal A}_{4},{\cal A}_{5}$ are related to all infinite t and $u'$ singularities of string amplitude appropriately. Also note that 
${\cal A}_{3},{\cal A}_{4}$ do carry $p.\xi_1$ singular terms , whose momenta are located in the bulk directions and from now on due to the presence of $p^i$, we are going to call them bulk singularities and start producing them in effective field theory (EFT) as well.

\section{An infinite $u'$ channel tachyon singularities}

In order to deal with all  singularities of S-matrix, we first need to have the entire expansion of Gamma functions. The expansion of $t L_3$ around  $t \rightarrow 0,  \quad  s \rightarrow  -\frac{1}{4} , \quad u \rightarrow  -\frac{1}{4}$ is given by

\beqa
t L_3&=&\pi^{3/2}\bigg(\frac{1}{u'}\sum_{n=-1}^{\infty}c_n(s'+t)^{n+1}+\sum_{p, n,m=0}^{\infty}f_{p,n,m}(u')^{p}(t s')^{n}(t+s')^m\bigg),\labell{high66}\eeqa
with some of the coefficients as
\beqa
&&c_{-1}=1,\,c_0=0,\,c_1=\frac{1}{6}\pi^2,f_{0,0,1}=\frac{1}{3}\pi^2,f_{1,0,1}=f_{0,0,2}=6\z(3).
\nonumber
\eeqa
\vskip.1in
 
 The results for the trace that include $\gamma^{11}$ can also be true for the following 
\beqa
  p>3 , H_n=*H_{10-n} , n\geq 5.
  \nonumber\eeqa

We first extract the trace in ${\cal A}_{5}$ and write down all tachyon singularities as follows
\beqa
&&(4ik_3.\xi_2)k_{1c}\xi_{1i}\frac{16}{(p+1)!}(\pi^{3/2})(\mu'_p\beta'\pi^{1/2}) H^{i}_{a_{0}\cdots a_{p-1}}\eps^{a_{0}\cdots a_{p-1}c}\nonumber\\&&\times
\sum_{n=-1}^{\infty}c_n\frac{1}{u'}(s'+t)^{n+1}\Tr(\lam_1\lam_2\lam_3)
\label{bbx}\eeqa
where we used $(\mu'_p\beta'\pi^{1/2})$ as normalization constant to the S-matrix. To reconstruct all infinite tachyon singularities, one has to consider the following sub amplitude in Field theory 
\beqa
{\cal A}&=&V^{\alpha}(C_{p},\phi_1,T)G^{\alpha\beta}(T)V^{\beta}(T,T_3,A_2).\labell{amp42}\eeqa

Note that tachyon kinetic term $2\pi\alpha' D^aT D_aT $ has already been fixed in the DBI action  and it has no correction, also 
$V^{\beta}(T,T_3,A_2)$ vertex operator comes from tachyon kinetic term by taking ($D_aT=\partial_aT-i[A^a,T]$) so it has no correction either. Both Taylor expansion and 
pull-back of $C_{a_{0}...a_{p-1}}$ are needed. A field strength in the string amplitude $H^{i}_{a_{0}...a_{p-1}}$
being reproduced by pull-back and Taylor expansion of the $C_{a_{0}...a_{p-1}} $.
%$V^{\alpha}(C_{p},\phi_1,T)$ should be read off from the mixing Chern-Simons and Taylor expansion as follows 
Consider the following coupling

 \beqa
    2i\beta'\mu'_p (2\pi\alpha')^2\int_{\Sigma_{p+1}} \Tr(\partial_{i}C_{p}\wedge DT\phi^i)
    \label{jj}\eeqa
    
Above coupling is the mixing Chern-Simons and Taylor expansion of scalar field. We need to take into account  
$2i\beta'\mu'_p (2\pi\alpha')^2  \int_{\Sigma_{p+1}} \epsilon^{a_{0}..a_{p}} C_{ia_{0}...a_{p-2}} D_{a_{p-1}} \phi^i D_{a_{p}}T$ coupling as well. If we take integration by parts $D_{a_{p-1}}$ can just act on C- field. Because of the $\epsilon$ 
tensor it gives zero result if it acts on $D_{a_{p}}T$. Now if we take into account  $\epsilon^{a_{0}..a_{p}} \partial_{[a_{p-1}} C_{a_{0}...a_{p-2}]i} \phi^i  D_{a_{p}}T$
where the bracket  corresponds to antisymmetrization of the corresponding world-volume indices and
use the definition $H^{i}_{a_{0}...a_{p-1}}=p\partial_{i} C_{a_{0}...a_{p-1}}+\partial_{[a_{p-1}} C_{a_{0}...a_{p-2}]i}$, then we clarify that
both terms are needed to reproduce the string amplitude term 
$H^{i}_{a_{0}...a_{p-1}} k_{a_{p}} \xi^{i}$.
\vskip.1in

Note that the contributions from Taylor expansion and the other coupling in  momentum space would be 
$\epsilon^{a_{0}..a_{p}} p_{i} C_{a_{0}...a_{p-1}} k_{a_{p}} \xi^i$ and  ($-\epsilon^{a_{0}..a_{p}} p_{a_{p-1}} C^{i}_{a_{0}...a_{p-2}} k_{a_{p}} \xi^i$)  accordingly, thus one obtains the following counterparts for the vertices in EFT

\beqa
V^{\beta}(T,T_3,A_2)&=&2T_p(2\pi\alpha')k_3.\xi_2
\Tr(\lam_2\lam_3\Lambda^\beta),\nonumber\\
V^{\alpha}(C_{p},\phi_1,T)&=&2\mu'_p\beta'\frac{(2\pi\alpha')^{2}}{(p+1)!}\epsilon^{a_0\cdots a_{p}}H^{i}_{a_0\cdots a_{p-1}}k_{a_p}\xi_{1i}\Tr(\lam_1\Lambda^{\alpha}),\nonumber\\
G^{\alpha\beta}(T) &=&\frac{-i\delta^{\alpha\beta}}{(2\pi\alpha') T_p
(k^2+m^2)}=\frac{-i\delta^{\alpha\beta}}{(2\pi\alpha') T_p
(u')}.
\label{moi}\eeqa

where $k$ in the above is momentum of off-shell tachyon $k=k_2+k_3=-(p+k_1)$. Now by replacing \reef{moi} inside  \reef{amp42} one obtains 
\beqa
&&(4ik_3.\xi_2)\xi_{1i}\frac{16}{(p+1)!}\pi^{2} \mu'_p\beta' H^{i}_{a_{0}\cdots a_{p-1}}(p+k_1)_{a_{p}}\eps^{a_{0}\cdots a_{p}}\frac{1}{u'}\label{bbx2}\eeqa

where if one uses the  identity  $p_{a_{p}}\eps^{a_{0}\cdots a_{p}}=0$\footnote{Note that, the derivation of the identity  $p_a \epsilon^{a_0...a_{p-1}a}=0$  can be  found from  various equations of \cite{Hatefi:2015gwa}, for instance it is clarified in formula (9) of \cite{Hatefi:2015gwa} , namely, to actually obtain the same result for the amplitude of a three point function of one RR and a scalar field in both  symmetric and asymmetric S-matrix , this identity  $p_a \epsilon^{a_0...a_{p-1}a}=0$ should be true. The other example to prove the above identity is as follows. We have just shown  it in  section 5 of \cite{Hatefi:2015gwa}, basically  to get to the same result for the S-matrix of  a four point function of $<C^{-1} T^0 \phi^{-1}>$ and $<C^{-2} T^0 \phi^{0}>$ , that identity should be employed  ( notice to the footnote  19 of \cite{Hatefi:2015gwa}). Finally, we  have shown that  if and only if that identity holds, then  certainly the S-matrix of  $<C^{-1} A^{-1} T^0 T^0>$ of  \cite{Hatefi:2012cp}  does satisfy Ward identity related to the gauge field.}, then one reveals that the first $u'$ channel tachyon pole of \reef{bbx} can be precisely produced. However, in \reef{bbx} we do have infinite poles  and  the only way to regenerate them is  to induce the higher derivative corrections as follows
\beqa
\frac{2i\beta'\mu_p'}{p!}(2\pi\alpha')^2 \partial_{i}C_{p}\wedge \Tr\left(\sum_{n=-1}^{\infty}c_n(\alpha')^{n+1}  D_{a_1}\cdots D_{a_{n+1}}DT D^{a_1}...D^{a_{n+1}}\phi^i\right) \labell{highaa}\eeqa
 
One needs to be reminded about the facts that the tachyon propagator and the vertex of two tachyon and one gauge field , do not receive any correction as they have been derived from the kinetic term of tachyons (as it has been already fixed) , that is why we claim that, taking \reef{highaa} is the only way of reconstructing all the singularities.

 Having set \reef{highaa}, we were able to actually obtain the extension of the vertex operator to all orders as follows 
\beqa
V^{\alpha}(C_{p},T,\phi_1)=\frac{2\mu'_p\beta'(2\pi\alpha')^{2} k_{a_p}\xi_{1i}}{(p)!}\epsilon^{a_0\cdots a_{p}}H^{i}_{a_0\cdots a_{p-1}}\sum_{m=-1}^{\infty}c_m(\alpha'k_1\cdot k)^{m+1}\Tr(\lam_1\Lambda^{\alpha}).\label{2300}\eeqa

Substituting \reef{2300} to \reef{amp42}, considering the fixed propagator (the fixed two tachyons, a gauge field vertex operator)  and making use of the following identity   $\sum_{m=-1}^{\infty}c_m(\alpha'k_1\cdot k)^{m+1}=\sum_{m=-1}^{\infty}c_m (t+s')^{m+1}$,  one exactly  reconstructs all  infinite $u'$ tachyon poles of \reef{bbx} in the effective field theory side as well. It is also worth pointing out the fact that by comparisons we get to know that there was no residual contact interactions to be left over in the EFT. Now we turn to infinite $t$ channel scalar singularities.

\vskip.1in

\section{An infinite $t$ channel scalar singularities}

Having expanded  $u' L_3$ around  the same $t \rightarrow 0,  \quad  s \rightarrow  -\frac{1}{4} , \quad u \rightarrow  -\frac{1}{4}$ , we get

 \beqa
u'L_3&=&\pi^{3/2}\bigg(\frac{1}{t}\sum_{n=-1}^{\infty}c_n(u'+s')^{n+1}
+\sum_{p,n,m=0}^{\infty}f_{p,n,m}t^{p}(u' s')^{n}(u'+s')^m\bigg),\label{tpoles}\eeqa
where  the coefficients of $c_n, f_{p,n,m}$ are read.\footnote{
\beqa
&& c_2=2\z(3), f_{2,0,0}=f_{0,1,0}=2\z(3),f_{1,0,0}=\frac{1}{6}\pi^2,f_{1,0,2}=\frac{19}{60}\pi^4,\nonumber\\
&&f_{0,0,1}=\frac{1}{3}\pi^2,f_{0,0,3}=f_{2,0,1}=\frac{19}{90}\pi^4,f_{1,1,0}=f_{0,1,1}=\frac{1}{30}\pi^4.\labell{865} \nonumber\eeqa.} After extracting the trace in ${\cal A}_{6}$ one writes  down all scalar t-channel singularities of the S-matrix as follows

\beqa
 4ik_1.\xi_2  k_{3c} \xi_{1i} \frac{16  \mu'_p\beta'\pi^{2}}{(p+1)!}\sum_{n=-1}^{\infty}c_n\frac{1}{t}(s'+u')^{n+1}
H^{i}_{a_{0}\cdots a_{p-1}}
\eps^{a_{0}\cdots a_{p-1}c }\Tr(\lam_1\lam_2\lam_3).\label{bbx32}\eeqa
Later on  we consider the restricted Bianchi identity to actually
write down $p_c H^{i}_{a_{0}\cdots a_{p-1}}\eps^{a_{0}\cdots a_{p-1}c }$ in terms of 
$p^i H_{a_{0}\cdots a_{p}}\eps^{a_{0}\cdots a_{p} }$and  generate  all the bulk singularities of the  S-matrix in EFT as well.  
 To construct all infinite  scalar singularities, one must  consider the following 
 sub amplitude in field theory 

 \beqa
{\cal A}&=&V^{\alpha}_i(C_{p},T_3,\phi)G^{\alpha\beta}_{ij}(\phi)V^{\beta}_j(\phi,\phi_1,A_2),\labell{amp4666}\eeqa

Note that the scalar kinetic term  has already been fixed in the DBI action  and it has no correction, also 
$V^{\beta}_j(\phi,\phi_1,A_2)$ can be derived just from scalar kinetic term by extracting the covariant 
derivative of scalar field ($D_a \phi^i=\partial_a \phi^i-i[A_a,\phi^i]$) so it has no correction either. 
%$V^{\alpha}_i(C_{p},\phi_1,T)$ has been read off from the mixing Chern-Simons and Taylor expansion as appeared in \reef{jj}, therefore 
One derives the following vertices in EFT 

\beqa
V^{\beta}_j(\phi,\phi_1,A_2)&=&-2 (2\pi\alpha')^2 T_p k_1.\xi_2 \xi_{1j} \Tr(\lam_1\lam_2\Lambda^{\beta}),\label{mmmpo}\eeqa
\beqa
V^{\alpha}_i(C_{p},\phi,T_3)&=&2\mu'_p\beta'\frac{(2\pi\alpha')^{2}}{(p+1)!}\epsilon^{a_0\cdots a_{p}}H^{i}_{a_0\cdots a_{p-1}}k_{3_{a_{p}}}\Tr(\lam_3\Lambda^{\alpha}),\nonumber\\
G^{\alpha\beta}_{ij}(\phi) &=&\frac{-i\delta^{\alpha\beta}\delta^{ij}}{(2\pi\alpha')^2 T_p
(k^2)}=\frac{-i\delta^{\alpha\beta}\delta^{ij}}{(2\pi\alpha')^2 T_p
(t)}.
\label{moi3}\eeqa

where $k_3$ in the above is momentum of on-shell tachyon and we used momentum conservation along 
the world volume direction as well. Now by replacing \reef{moi3} inside  \reef{amp4666} one 
concludes that the first t-channel scalar pole of 
 \reef{bbx32} can be precisely produced. However, as it is clear in \reef{bbx32} we do have  infinite poles  and  the only way to reproduce them is  to propose the higher derivative 
 corrections to the  actions.

Taking  \reef{highaa} into account one can get  the  all order extensions of the vertex operator  as 

\beqa
V^{\alpha}_{i}(C_{p},\phi,T_3)&=&2\mu'_p\beta'\frac{(2\pi\alpha')^{2}}{(p+1)!}
\epsilon^{a_0\cdots a_{p}}H^{i}_{a_0\cdots a_{p-1}} k_{3_{a_p}}\sum_{m=-1}^{\infty}c_m(\alpha'k_3\cdot k)^{m+1}
\Tr(\lam_3\Lambda^{\alpha}),\label{23061}\eeqa

\vskip.1in

Using momentum conservation, one gets,  $(\alpha'k_3\cdot k)= (u'+s')$. 
If we  substitute \reef{23061} inside   \reef{amp4666} and  keeping 
fixed all the other vertices , one clarifies the fact that all the 
infinite t-channel poles of \reef{bbx32} are precisely gained in the effective field 
theory side as well, so the higher derivative corrections of \reef{highaa} are exact.

 \vskip.1in
 
 To generate all the contact interactions, we just highlight the following references  \cite{Hatefi:2012wj,Hatefi:2012rx}.
 Also notice to the point that for this S-matrix we do have external gauge field as well as an external scalar and a
 real tachyon therefore  in the  action of \reef{jj}, one needs to first  consider the presence of commutator inside  the covariant derivative of tachyon, so that external gauge field shows up and then try to apply higher derivative corrections properly. Lets turn to the main point of the paper which is indeed dealing 
with all the bulk singularity structures of the S-matrix.

\section{An Infinite $u'$  Bulk Singularity Structures }

Apart from an infinite number of $u'$ channel tachyon poles , we have an infinite number of bulk singularity structures that can be accommodated in EFT as well. 

Consider the expansion of  \reef{high66} and do replace it into the entire  ${\cal A}_{4}$ of \reef{48} , extract the related trace as well as normalize this fourth part of S-matrix so that the whole bulk singularities are now found out as follows 
 
\beqa
&&  p.\xi_1 (-4ik_3.\xi_2) \frac{16 \mu'_p\beta'\pi^{2} }{(p+1)!}  H_{a_{0}\cdots a_{p}}\eps^{a_{0}\cdots a_{p}} 
\sum_{n=-1}^{\infty}c_n\frac{1}{u'}(s'+t)^{n+1}\Tr(\lam_1\lam_2\lam_3)
\label{bbx412}\eeqa

We call them bulk singularities, because they do involve  all the infinite momenta of RR in the bulk directions (due to an infinite number of $p.\xi_1$ terms). In the other words, we can keep track of these terms that carry the scalar product of momentum of RR in the bulk and scalar polarisation  and claim that these bulk singularities  are needed in the entire S-matrix as we are going to reconstruct them in EFT as well.

To reconstruct all these infinite bulk $u'$ channel  singularities, one has to once more, consider the following sub amplitude in field theory side 
\beqa
{\cal A}&=&V^{\alpha}(C_{p},\phi_1,T)G^{\alpha\beta}(T)V^{\beta}(T,T_3,A_2).\labell{amp4298}\eeqa

It is worth highlighting the remark that both tachyon propagator and $V^{\beta}(T,T_3,A_2)$ vertex operator, will not receive any correction.

To actually produce $p.\xi_1$ terms in EFT one needs to  consider the following integrations
%\reef{jj} and  take integration as well , so that  
 \beqa
    2i\beta'\mu'_p (2\pi\alpha')^2\int_{\Sigma_{p+1}} \bigg(-\Tr(\partial_{i}  d_{a_{p}} C_{a_{0}...a_{p-1}}  T\phi^i)- \Tr(\partial_{i} C_{a_{0}...a_{p-1}}  T d_{a_{p}} \phi^i)\bigg),
    \label{jj56}\eeqa

where we suppose all the fields are zero at infinity. 
%Indeed in \reef{moi}, 
We have already taken into account  the contribution of the second term of the above action  and were able to produce all infinite $u'$ channel tachyon singularities, so here we just need to consider the contribution of the first term of \reef{jj56} to actually derive the following vertex operator 

\beqa
%V^{\beta}(T,T_3,A_2)&=&2T_p(2\pi\alpha')k_3.\xi_2\Tr(\lam_2\lam_3\Lambda^\beta),\nonumber\\
V^{\alpha}(C_{p},\phi_1,T)&=&2\mu'_p\beta'\frac{(2\pi\alpha')^{2}}{(p+1)!}\epsilon^{a_0\cdots a_{p}}H_{a_0\cdots a_{p}}p.\xi_{1}\Tr(\lam_1\Lambda^{\alpha})
%G^{\alpha\beta}(T) &=&\frac{-i\delta^{\alpha\beta}}{(2\pi\alpha') T_p(k^2+m^2)}=\frac{-i\delta^{\alpha\beta}}{(2\pi\alpha') T_p(u')}.
\label{moi881}\eeqa
Keeping fixed   $V^{\beta}(T,T_3,A_2)$ and the tachyon propagator and  replacing \reef{moi881} inside  \reef{amp4298} one gets 
  \beqa
&&(-4ik_3.\xi_2) p.\xi_1\frac{16 \mu'_p\beta'\pi^{2} }{u' (p+1)!}  H_{a_{0}\cdots a_{p}}\eps^{a_{0}\cdots a_{p}} \Tr(\lam_1\lam_2\lam_3)
\label{bbx41243}\eeqa
 
  One can observe the fact that \reef{bbx41243} is indeed the first bulk singularity of the S-matrix (consider $n=-1$ inside \reef {bbx412}). To actually even produce all infinite bulk singularities , one needs to apply all higher derivative corrections to the first part of the above action as follows 
  
\beqa
\frac{-2i\beta'\mu_p'}{(p+1)!}(2\pi\alpha')^2 \int_{\Sigma_{p+1}}\bigg(\Tr \sum_{n=-1}^{\infty}c_n(\alpha')^{n+1} \partial_{i}  d_{a_{p}} C_{a_{0}...a_{p-1}}  D_{a_1}\cdots D_{a_{n+1}}T D^{a_1}...D^{a_{n+1}}\phi^i\bigg) \labell{highaa9811}\eeqa
 
 Having set \reef{highaa9811}, we are able to precisely reconstruct all order extensions of the above vertex operator  as follows 
\beqa
V^{\alpha}(C_{p},T,\phi_1)=\frac{2\mu'_p\beta'(2\pi\alpha')^{2} p.\xi_{1}}{(p+1)!}\epsilon^{a_0\cdots a_{p}}H_{a_0\cdots a_{p}}\sum_{m=-1}^{\infty}c_m(t+s')^{m+1}\Tr(\lam_1\Lambda^{\alpha}).\label{2322}\eeqa

Substituting \reef{2322} to \reef{amp4298}, considering the fixed propagator 
(and fixed two tachyons, a gauge field vertex operator), one is able to exactly regenerate  all  infinite $u'$ bulk poles of \reef{bbx412} in the effective field theory side as well. Eventually in the next section , we try  to produce an infinite $t$ channel bulk Singularity structures in the effective field theory side too. 
\section{An Infinite $t$ Channel Bulk Singularity structures}

Apart from an infinite number of $t$ channel scalar poles , we do have an infinite number of $t$ channel bulk singularity structures that can be found out in EFT as well. 

Consider the expansion of  \reef{tpoles} and substitute it into the whole ${\cal A}_{3}$ of \reef{48}, extract the  trace as well as normalize this third part of S-matrix, to indeed get to the  whole bulk t- channel singularities from the string theory point of view as follows
 
\beqa
&&  p.\xi_1 (4ik_1.\xi_2) \frac{16 \mu'_p\beta'\pi^{2} }{(p+1)!}  H_{a_{0}\cdots a_{p}}\eps^{a_{0}\cdots a_{p}} 
\sum_{n=-1}^{\infty}c_n\frac{1}{t}(s'+u')^{n+1}\Tr(\lam_1\lam_2\lam_3)
\label{bbx4123}\eeqa

These are also bulk singularities, because  all the infinite singularity terms that involve momentum of RR in the bulk have been embedded into the S-matrix and in below we want to show that they even can be  reconstructed  in EFT as well. To regenerate all these infinite bulk $t$ channel  scalar singularities, one needs to deal with the following sub amplitude in field theory side 
 \beqa
{\cal A}&=&V^{\alpha}_i(C_{p},T_3,\phi)G^{\alpha\beta}_{ij}(\phi)V^{\beta}_j(\phi,\phi_1,A_2),\labell{amp466622}\eeqa
Note that  both scalar propagator and $V^{\beta}_j(\phi,\phi_1,A_2)$ vertex operator, do not get corrected.
To actually produce $p.\xi_1$ terms in EFT one should make use of Taylor expansion
%\reef{jj} 
and  take integration by parts as well.
%as we did in \reef{jj56}. 
 Indeed in \reef{moi3}, we have considered the contribution of  the second term of \reef{jj56} and precisely produced all infinite $t$ channel scalar  singularities of ${\cal A}_{6}$ , in the meantime,  here we just need to consider the contribution from the first term of \reef{jj56} to actually explore 
\beqa
V^{\alpha}_{i}(C_{p},\phi,T_3)&=&-2p_i\mu'_p\beta'\frac{(2\pi\alpha')^{2}}{(p+1)!}
\epsilon^{a_0\cdots a_{p}}H_{a_0\cdots a_{p}} \Tr(\lam_3\Lambda^{\alpha})\label{moi88122}\eeqa

Keeping fixed   $V^{\beta}_j(\phi,\phi_1,A_2)$ and the scalar propagator and  replacing \reef{moi88122} inside  \reef{amp466622} one gains 
  \beqa
&&(4ik_1.\xi_2) p.\xi_1\frac{16 \mu'_p\beta'\pi^{2} }{t(p+1)!}  H_{a_{0}\cdots a_{p}}\eps^{a_{0}\cdots a_{p}} \Tr(\lam_1\lam_2\lam_3)
\label{bbx412367}\eeqa       
  One can observe the fact that \reef{bbx412367} is indeed the first bulk singularity of the S-matrix (consider $n=-1$ inside \reef {bbx4123}). To be able to produce all infinite bulk singularities , one needs to apply all higher derivative corrections properly.
  % as we did in\reef{highaa9811}.
Having set \reef{highaa9811}, we are able to precisely reconstruct all order extensions of $V^{\alpha}_{i}(C_{p},\phi,T_3)$   vertex operator  as follows 
\beqa
V^{\alpha}_{i}(C_{p},\phi,T_3)&=&-2p_i \mu'_p\beta'\frac{(2\pi\alpha')^{2}}{(p+1)!}\epsilon^{a_0\cdots a_{p}}H_{a_0\cdots a_{p}} \sum_{m=-1}^{\infty}c_m(u'+s')^{m+1}\Tr(\lam_3\Lambda^{\alpha}).\label{230612}\eeqa

Substituting \reef{230612} to \reef{amp466622}, considering the fixed scalar propagator (and the fixed two scalars, a gauge field vertex operator),  one is able to precisely regenerate  all  infinite $t$ channel bulk  singularities of \reef{bbx4123} in the effective field theory side as well. Therefore, we were able to even produce all the infinite bulk singularities $u',t$ in the EFT and that evidently confirms that the presence of bulk  singularities is needed inside the entire S-matrix. Note that, it also has the important  consequences for which  one of them is the essential appearances of new restricted   Bianchi identities on world volume directions of D-branes that we got in this particular  S-matrix.  

\section*{Acknowledgments}

The author would like to thank  C. Hull, A. Tseytlin and D. Waldram for useful discussions  at Imperial College in London, he also thanks N. Lambert for having several valuable discussions at Kings College London. He warmly thanks P. Vanhove, T. Damour, S. Shatashvili, M. Kontsevich and V. Pestun at IHES and his colleagues at QMUL. In particular he thanks C. Vafa, R. Russo for various discussions ,  C. Papageorgakis, S.Thomas , D. Young, R.Russo, A.Brandhuber, G.Travaglini, D. Berman, S. Ramgoolam, A. Sen , R.Myers, K. Narain, L.Alvarez-Gaume, W. Lerche, J.Polchinski, N. Arkani-Hamed,  W. Siegel, E.Witten, H. Steinacker, A. Rebhan,D. Grumiller ,T. Wrase, and P. Anastasopoulos for  valuable discussions. Some parts of this work have been done at Caltech, Simons Centre , UC Berkeley , IAS at Princeton, Harvard, CERN and  IHES. He is very grateful to those institutes for providing such an exciting and challenging environment. This work was supported by the FWF project P26731-N27.

\end{document}